\documentstyle[12pt,epsfig,graphics]{article}
\begin{document}\bibliographystyle{plain}\begin{titlepage}
\renewcommand{\thefootnote}{\fnsymbol{footnote}}\hfill\begin{tabular}{l}
HEPHY-PUB 734/00\\UWThPh-2000-42\\hep-ph/0011235\\October 2000\end{tabular}
\\[1cm]\Large\begin{center}{\bf INSTANTANEOUS BETHE--SALPETER EQUATION:
ANALYTIC APPROACH FOR NONVANISHING MASSES OF THE BOUND-STATE CONSTITUENTS}\\
\vspace{0.8cm}\large{\bf Wolfgang LUCHA\footnote[1]{\normalsize\ {\em E-mail
address\/}: wolfgang.lucha@oeaw.ac.at}}\\[.3cm]\normalsize Institut f\"ur
Hochenergiephysik,\\\"Osterreichische Akademie der
Wissenschaften,\\Nikolsdorfergasse 18, A-1050 Wien,
Austria\\[0.7cm]\large{\bf Khin MAUNG MAUNG\footnote[3]{\normalsize\ {\em
E-mail address\/}: maung@jlab.org}}\\[.3cm]\normalsize Department of Physics,
Hampton University,\\Hampton, VA 23668\\[0.7cm]\large{\bf Franz F.
SCH\"OBERL\footnote[2]{\normalsize\ {\em E-mail address\/}:
franz.schoeberl@univie.ac.at}}\\[.3cm]\normalsize Institut f\"ur Theoretische
Physik, Universit\"at Wien,\\Boltzmanngasse 5, A-1090 Wien, Austria\vfill
{\normalsize\bf Abstract}\end{center}\normalsize The instantaneous
Bethe--Salpeter equation, derived from the general Bethe--Salpeter formalism
by assuming that the involved interaction kernel is instantaneous, represents
the most promising framework for the description of hadrons as bound states
of quarks from first quantum-field-theoretic principles, that is, quantum
chromodynamics. Here, by extending a previous analysis confined to the case
of bound-state constituents with vanishing masses, we demonstrate that the
instantaneous Bethe--Salpeter equation for bound-state constituents with
(definitely) nonvanishing masses may be converted into an eigenvalue problem
for an explicitly---more precisely, algebraically---known matrix, at least,
for a rather wide class of interactions between these bound-state
constituents. The advantages of the explicit knowledge of this matrix
representation are self-evident.\vspace{3ex}

\renewcommand{\thefootnote}{\arabic{footnote}}\end{titlepage}

\normalsize

\section{Introduction}In principle, the appropriate tool for the description
of bound states within relativistic quantum field theory is the
Bethe--Salpeter formalism. However, attempts to solve~the Bethe--Salpeter
equation face several, well-known obstacles. In view of this, in practical
applications usually the easier-to-handle instantaneous approximation for the
involved interaction kernel is considered.

In a recent paper \cite{Lucha00:IBSEm0}, we introduced, for the somewhat
simpler example of massless bound-state constituents, a technique for
converting the instantaneous Bethe--Salpeter equation into an eigenvalue
problem for an explicitly given matrix. Here, the analysis of
Ref.~\cite{Lucha00:IBSEm0} is extended to the case of nonvanishing masses of
the bound-state constituents.

\section{Bethe--Salpeter Equation as a Matrix Equation}Our goal is to
demonstrate the possibility of converting the Bethe--Salpeter equation~to a
matrix equation. Thus, let us accept the same simplifying assumptions as in
Ref.~\cite{Lucha00:IBSEm0}:\begin{itemize}\item The propagators in the
Bethe--Salpeter equation may be approximated by free propagators with some
kind of effective masses of the bound-state constituents.\item The
bound-state constituents have equal masses.\end{itemize}

Paralleling the discussion in Ref.~\cite{Lucha00:IBSEm0}, we consider
fermion--antifermion bound states with spin $J,$ parity $P=(-1)^{J+1}$ and
charge-conjugation quantum number $C=(-1)^J.$ The corresponding equal-time
Bethe--Salpeter amplitude, or ``Salpeter amplitude,'' $\chi$ involves two
independent components, $\Psi_1$ and~$\Psi_2.$ For two fermions of equal
masses~$m$ and internal momentum~$\bf k,$ in the bound state's rest frame it
reads in momentum~space$$\chi({\bf k})=\left[\Psi_1({\bf
k})\,\frac{m-\mbox{\boldmath{$\gamma$}}\cdot{\bf k}}{E(k)}+\Psi_2({\bf
k})\,\gamma^0\right]\gamma_5\ ;$$$E(k)\equiv\sqrt{k^2+m^2},$ $k\equiv|{\bf
k}|,$ is the energy of a free particle of mass $m$ and momentum~${\bf k}.$ We
confine ourselves to the case $J=0,$ that is, to bound states with
spin-parity-charge conjugation assignment $J^{PC}=0^{-+},$ denoted by
${}^1{\rm S}_0$ in usual spectroscopic notation.

By expanding the Salpeter amplitude into some convenient set of basis
matrices in Dirac space and after factorizing off the (vector) spherical
harmonics depending on the angular variables, the instantaneous
Bethe--Salpeter equation for fermion--antifermion bound states can be reduced
to a set of coupled equations for radial wave functions~\cite{Lagae92}. For
pure time-component Lorentz-vector interactions, i.e.,
$\gamma^0\otimes\gamma^0$ kernels, it reads~\cite{Lagae92,Olsson95}
\begin{eqnarray} 
&&2\,E(k)\,\Psi_2(k)+\int\limits_0^\infty\frac{{\rm d}k'\,k'^2}{(2\pi)^2}
\,V_0(k,k')\,\Psi_2(k')=M\,\Psi_1(k)\
,\nonumber\\[1ex]&&2\,E(k)\,\Psi_1(k)\label{Eq:IBSE}\\[1ex]&&
+\int\limits_0^\infty\frac{{\rm d}k'\,k'^2}{(2\pi)^2}
\left[\frac{m}{E(k)}\,V_0(k,k')\,\frac{m}{E(k')}
+\frac{k}{E(k)}\,V_1(k,k')\,\frac{k'}{E(k')}\right]\Psi_1(k')=M\,\Psi_2(k)\
,\nonumber\end{eqnarray}where, expressed in terms of a static interaction
potential $V(r)$ in configuration space,$$V_L(k,k')\equiv
8\pi\int\limits_0^\infty{\rm d}r\,r^2\,V(r)\,j_L(k\,r)\,j_L(k'\,r)\ ,\quad
L=0,1,2,\dots\ .$$Here, $j_n(z)$ ($n=0,\pm 1,\pm2,\dots$) are the spherical
Bessel functions of the first kind \cite{Abramowitz}.

Expressing, for $M\ne0,$ from the first of Eqs.~(\ref{Eq:IBSE}) the component
$\Psi_1$ in terms~of~the component $\Psi_2$ of the Salpeter amplitude $\chi$
and inserting this into the second of Eqs.~(\ref{Eq:IBSE}) yields an
eigenvalue equation for $\Psi_2$ with the bound-state mass squared as
eigenvalue:\begin{eqnarray*}M^2\,\Psi_2(k)&=&4\,E^2(k)\,\Psi_2(k)
+2\,E(k)\int\limits_0^\infty\frac{{\rm d}k'\,k'^2}{(2\pi)^2}\,
V_0(k,k')\,\Psi_2(k')\\[1ex]&+&2\int\limits_0^\infty\frac{{\rm
d}k'\,k'^2}{(2\pi)^2}\left[\frac{m^2}{E(k)}\,V_0(k,k')
+\frac{k\,k'}{E(k)}\,V_1(k,k')\right]\Psi_2(k')\\[1ex]
&+&\int\limits_0^\infty\frac{{\rm d}k'\,k'^2}{(2\pi)^2}
\left[\frac{m}{E(k)}\,V_0(k,k')\,\frac{m}{E(k')}
+\frac{k}{E(k)}\,V_1(k,k')\,\frac{k'}{E(k')}\right]\\[1ex]
&\times&\int\limits_0^\infty\frac{{\rm d}k''\,k''^2}{(2\pi)^2}\,V_0(k',k'')\,
\Psi_2(k'')\ .\end{eqnarray*}

In order to convert this eigenvalue equation to matrix form, we introduce
suitable sets of basis functions for the Hilbert space $L_2(R^+)$ of (with
weight function $w(x)=x^2$) square-integrable functions $f(x)$ on the
positive real line $R^+$ (see \cite{Lucha00:IBSEm0}). For a given value
$\ell=0,1,2,\dots$ of the angular momentum of their counterparts in three
dimensions,~the basis functions are called $\phi_i^{(\ell)}(r)$ in
configuration space and $\phi_i^{(\ell)}(p)$ in momentum~space. They are the
same as in Ref.~\cite{Lucha00:IBSEm0} apart from the fact that the real
variational parameter $\mu>0$ there is replaced here by the mass $m$ of the
bound-state constituents and~that, consequently, normalizability of the basis
vectors demands $m>0;$ their main features are summarized in
Appendix~\ref{App:Laguerre-nzm}. Expanding $\Psi_2(p)$ in terms of the radial
basis functions $\phi_i^{(0)}(p),$ the instantaneous Bethe--Salpeter equation
is solved by diagonalizing a matrix:$${\cal M}_{ij}=A_{ij}+B_{ij}
+C^{(1)}_{ij}+C^{(2)}_{ij}+D^{(1)}_{ij}+D^{(2)}_{ij}\ ,$$with the
abbreviations\begin{eqnarray*}A_{ij}&\equiv&4\int\limits_0^\infty{\rm
d}k\,k^2\,E^2(k)\,\phi_i^{(0)}(k)\,\phi_j^{(0)}(k)\ ,\\[1ex]
B_{ij}&\equiv&\frac{2}{(2\pi)^2}
\int\limits_0^\infty{\rm d}k\,k^2\,E(k)\,\phi_i^{(0)}(k)
\int\limits_0^\infty{\rm d}k'\,k'^2\,V_0(k,k')\,
\phi_j^{(0)}(k')\ ,\\[1ex] C^{(1)}_{ij}&\equiv&\frac{2\,m^2}{(2\pi)^2}
\int\limits_0^\infty\frac{{\rm d}k\,k^2}{E(k)}\,\phi_i^{(0)}(k)
\int\limits_0^\infty{\rm d}k'\,k'^2\,V_0(k,k')\,
\phi_j^{(0)}(k')\ ,\\[1ex] C^{(2)}_{ij}&\equiv&\frac{2}{(2\pi)^2}
\int\limits_0^\infty\frac{{\rm d}k\,k^3}{E(k)}\,\phi_i^{(0)}(k)
\int\limits_0^\infty{\rm d}k'\,k'^3\,V_1(k,k')\,\phi_j^{(0)}(k')\ ,\\[1ex]
D^{(1)}_{ij}&\equiv&\frac{m^2}{(2\pi)^4}
\int\limits_0^\infty\frac{{\rm d}k\,k^2}{E(k)}\,\phi_i^{(0)}(k)
\int\limits_0^\infty\frac{{\rm d}k'\,k'^2}{E(k')}\,V_0(k,k')
\int\limits_0^\infty{\rm d}k''\,k''^2\,V_0(k',k'')\,\phi_j^{(0)}(k'')\ ,\\[1ex]
D^{(2)}_{ij}&\equiv&\frac{1}{(2\pi)^4}
\int\limits_0^\infty\frac{{\rm d}k\,k^3}{E(k)}\,\phi_i^{(0)}(k)
\int\limits_0^\infty\frac{{\rm d}k'\,k'^3}{E(k')}\,V_1(k,k')
\int\limits_0^\infty{\rm d}k''\,k''^2\,V_0(k',k'')\,\phi_j^{(0)}(k'')\
.\end{eqnarray*}

The solution of this eigenvalue problem (of course, only for a finite matrix
size~$d$) proceeds, step by step, exactly along the lines presented in much
more detail in Ref.~\cite{Lucha00:IBSEm0}.

First of all, we introduce the matrix elements of the square $E^2$ of the
kinetic~energy:$$K_{ij}(m)\equiv\int\limits_0^\infty{\rm
d}k\,k^{2}\,E^2(k)\,\phi_i^{(0)}(k)\,\phi_j^{(0)}(k)\ .$$

Furthermore, in order to evaluate the terms
$B_{ij},C^{(1)}_{ij},\dots,D^{(2)}_{ij}$ analytically---which requires
repeated applications of the Fourier--Bessel transformations
(\ref{Eq:FB})---, we have~to expand several expressions involved in these
integrals in terms of the appropriate set of momentum-space basis functions
$\phi_i^{(\ell)}(k),$ $\ell=0,1$ (cf.\ Eqs.~(28), (31), (32) of
Ref.~\cite{Lucha00:IBSEm0}):\begin{eqnarray*}
E(k)\,\phi_i^{(0)}(k)&=&\sum_{j=0}^N\,b_{ji}(m)\,\phi_j^{(0)}(k)\ ,\\[1ex]
\frac{k}{E(k)}\,\phi_i^{(0)}(k)&=&\sum_{j=0}^N\,c_{ji}\,\phi_j^{(1)}(k)\
,\\[1ex] k\,\phi_i^{(0)}(k)&=&\sum_{j=0}^N\,d_{ji}(m)\,\phi_j^{(1)}(k)\
,\\[1ex]
\frac{1}{E(k)}\,\phi_i^{(0)}(k)&=&\sum_{j=0}^N\,e_{ji}(m)\,\phi_j^{(0)}(k)\ .
\end{eqnarray*}As consequence of the orthonormality of the momentum-space
basis functions~$\phi_i^{(\ell)}(p),$ the expansion coefficients $b_{ij}(m),$
$c_{ij},$ $d_{ij}(m),$ and $e_{ij}(m)$ may be expressed in the form (cf.\
Eqs.~(29), (33), (34) of Ref.~\cite{Lucha00:IBSEm0})\begin{eqnarray*}
b_{ij}(m)&=&\int\limits_0^\infty{\rm d}k\,k^2\,E(k)\,
\phi_i^{(0)}(k)\,\phi_j^{(0)}(k)\ ,\\[1ex]
c_{ij}&=&\int\limits_0^\infty\frac{{\rm d}k\,k^3}{E(k)}\,
\phi_i^{\ast(1)}(k)\,\phi_j^{(0)}(k)\ ,\\[1ex]
d_{ij}(m)&=&\int\limits_0^\infty{\rm d}k\,k^3\,
\phi_i^{\ast(1)}(k)\,\phi_j^{(0)}(k)\ ,\\[1ex]
e_{ij}(m)&=&\int\limits_0^\infty\frac{{\rm d}k\,k^2}{E(k)}\,
\phi_i^{(0)}(k)\,\phi_j^{(0)}(k)\ .\end{eqnarray*}These expansion
coefficients are, of course, not independent but satisfy several (only~in the
limit $N\to\infty,$ exact) relations of the kind\begin{eqnarray*}
&&\sum_{r=0}^N\,b_{ri}(m)\,b_{rj}(m)
=\sum_{r=0}^N\,d^\ast_{ri}(m)\,d_{rj}(m)+m^2\,\delta_{ij}=K_{ij}(m)\ ,\\[1ex]
&&\sum_{r=0}^N\,b_{ri}(m)\,e_{rj}(m)=\delta_{ij}\ ,\\[1ex]
&&\sum_{r=0}^N\,d_{ir}(m)\,e_{rj}(m)=c_{ij}\ .\end{eqnarray*}Employing these
relations in order to investigate systematically the errors induced by the
truncations of the expansion series in ${\cal M}_{ij},$ one finds that, for
instance, for $d=15$ (i.e., $15\times 15$ matrices) and for $N=49$ (i.e., a
truncation to the first 50 basis vectors), all the above relations are
satisfied with relative errors less than~$3\%.$

Finally, we'll need the expansions of the expressions
$V(r)\,\phi_i^{(\ell)}(r)$ in terms of
$\phi_i^{(\ell)}(r)$:$$V(r)\,\phi_i^{(\ell)}(r)=
\sum_{j=0}^N\,V^{(\ell)}_{ji}(m)\,\phi_j^{(\ell)}(r)\ ,\quad\ell=0,1\ .$$
Quite obviously, here $V^{(\ell)}_{ij}(m)$ is the real and symmetric matrix
of expectation values~of the interaction potential $V(r)$ with respect to our
basis functions $\phi_i^{(\ell)}(r)$ for a given $\ell$:\begin{equation}
V^{(\ell)}_{ij}(m)=\int\limits_0^\infty{\rm d}r\,r^2\,V(r)\,\phi_i^{(\ell)}(r)
\,\phi_j^{(\ell)}(r)\ .\label{Eq:EV-IP}\end{equation}In
Refs.~\cite{Lucha97,Lucha98O,Lucha98D} it has been shown that, for
interaction potentials of the power-law~form$$V(r)=\sum_na_n\, r^{b_n}$$(with
sets of arbitrary real constants $a_n$ and $b_n$), the expectation values
$V^{(\ell)}_{ij}(m)$ can~be easily worked out algebraically; for their
algebraic expression for the most general~case, consult either Sec.~4 of
Ref.~\cite{Lucha97}, or Sec.~3.10 of Ref.~\cite{Lucha98O}, or Sec.~2.8.1 of
Ref.~\cite{Lucha98D}.

With the aid of the above series expansions, the Fourier--Bessel
transformations~(\ref{Eq:FB}), and the definition (\ref{Eq:EV-IP}) of the
matrix elements $V^{(\ell)}_{ij}(m)$ of the interaction potential $V(r),$ the
matrix ${\cal M}_{ij}$ is approximated by the (at least, for all power-law
potentials) algebraic expression\begin{eqnarray}{\cal M}_{ij}&=&4\,K_{ij}(m)
+2\,\sum_{r=0}^N\,b_{ri}(m)\,V^{(0)}_{rj}(m)
+2\,m^2\,\sum_{r=0}^N\,e_{ri}(m)\,V^{(0)}_{rj}(m)\nonumber\\[1ex]
&+&2\,\sum_{r=0}^N\,\sum_{s=0}^N\,c^\ast_{ri}\,V^{(1)}_{rs}(m)\,d_{sj}(m)
+m^2\,\sum_{r=0}^N\,\sum_{s=0}^N\,\sum_{t=0}^N\,e_{ri}(m)\,V^{(0)}_{rs}(m)\,
e_{st}(m)\,V^{(0)}_{tj}(m)\nonumber\\[1ex]
&+&\sum_{r=0}^N\,\sum_{s=0}^N\,\sum_{t=0}^N\,c^\ast_{ri}\,
V^{(1)}_{sr}(m)\,c_{st}\,V^{(0)}_{tj}(m)\ .\label{Eq:M2-nzm}\end{eqnarray}

The explicit evaluation of the matrix elements $K_{ij}(m)$ of the kinetic
energy squared and of the various expansion coefficients $b_{ij}(m),$
$c_{ij},$ $d_{ij}(m),$ and $e_{ij}(m)$ is a tedious~but straightforward task.
One obtains, for $K_{ij}(m)$ (cf.\ Sec.~2.2 of Ref.~\cite{Lucha00:IBSEm0}),
\begin{eqnarray*}
K_{ij}(m)&=&\frac{4\,m^2}{\pi\,\sqrt{(i+1)\,(i+2)\,(j+1)\,(j+2)}}
\\[1ex]&\times&\sum_{r=0}^i\,\sum_{s=0}^j\,(-2)^{r+s}
\left(\begin{array}{c}i+2\\i-r\end{array}\right)
\left(\begin{array}{c}j+2\\j-s\end{array}\right)(r+1)\,(s+1)\\[1ex]&\times&
\left[\sum_{k=0}^{|r-s|}\left(\begin{array}{c}|r-s|\\k\end{array}\right)
\frac{\Gamma(\frac{1}{2}\,(k+1))\,\Gamma(\frac{1}{2}\,(1+r+s+|r-s|-k))}
{\Gamma(\frac{1}{2}\,(2+r+s+|r-s|))}
\cos\left(\frac{k\,\pi}{2}\right)\right.\\[1ex]
&-&\left.\sum_{k=0}^{r+s+4}\left(\begin{array}{c}r+s+4\\k\end{array}\right)
\frac{\Gamma(\frac{1}{2}\,(k+1))\,\Gamma(\frac{1}{2}\,(5+2\,r+2\,s-k))}
{\Gamma(3+r+s)}\cos\left(\frac{k\,\pi}{2}\right)\right],\end{eqnarray*}or, in
matrix form,$$K(m)\equiv\left(K_{ij}(m)\right)=
2\,m^2\left(\begin{array}{ccc}1&\displaystyle\frac{1}{\sqrt{3}}&\cdots\\[3ex]
\displaystyle\frac{1}{\sqrt{3}}&\displaystyle\frac{5}{3}&\cdots\\[3ex]
\vdots&\vdots&\ddots\end{array}\right),$$and, for the expansion coefficients
$b_{ij}(m),$\begin{eqnarray*}b_{ij}(m)
&=&\frac{4\,m}{\pi\,\sqrt{(i+1)\,(i+2)\,(j+1)\,(j+2)}}\\[1ex]
&\times&\sum_{r=0}^i\,\sum_{s=0}^j\,(-2)^{r+s}
\left(\begin{array}{c}i+2\\i-r\end{array}\right)
\left(\begin{array}{c}j+2\\j-s\end{array}\right)(r+1)\,(s+1)\\[1ex]&\times&
\left[\sum_{k=0}^{|r-s|}\left(\begin{array}{c}|r-s|\\k\end{array}\right)
\frac{\Gamma(\frac{1}{2}\,(k+1))\,\Gamma(\frac{1}{2}\,(2+r+s+|r-s|-k))}
{\Gamma(\frac{1}{2}\,(3+r+s+|r-s|))}
\cos\left(\frac{k\,\pi}{2}\right)\right.\\[1ex]
&-&\left.\sum_{k=0}^{r+s+4}\left(\begin{array}{c}r+s+4\\k\end{array}\right)
\frac{\Gamma(\frac{1}{2}\,(k+1))\,\Gamma(\frac{1}{2}\,(6+2\,r+2\,s-k))}
{\Gamma(\frac{1}{2}\,(7+2\,r+2\,s))}
\cos\left(\frac{k\,\pi}{2}\right)\right]\end{eqnarray*}or, in matrix form,
$$b(m)\equiv(b_{ij}(m))=\frac{64\,m}{5\pi}\left(\begin{array}{ccc}
\displaystyle\frac{1}{3}&\displaystyle\frac{1}{7\,\sqrt{3}}&\cdots\\[3ex]
\displaystyle\frac{1}{7\,\sqrt{3}}&\displaystyle\frac{11}{27}&\cdots\\[3ex]
\vdots&\vdots&\ddots\end{array}\right).$$In order to evaluate the expansion
coefficients $c_{ij}$ and $e_{ij}(m),$ we introduce the integrals
\begin{eqnarray}
I^{(n)}_{ij}(m)&\equiv&\int\limits_0^\infty\frac{{\rm d}k\,k^{2+n}}{E(k)}\,
\phi_i^{(0)}(k)\,\phi_j^{(0)}(k)\ ,\quad n=0,1,2,\dots\ ,\nonumber\\[1ex]
J^{(n)}_{ij}(m)&\equiv&\int\limits_0^\infty\frac{{\rm d}k\,k^{2+n}}{E(k)}\,
\phi_i^{\ast(1)}(k)\,\phi_j^{(0)}(k)\ ,\quad n=0,1,2,\dots\ ;
\label{Eq:integrals}\end{eqnarray}the (somewhat lengthy) explicit expressions
of these integrals are given in Appendix~\ref{App:integrals}. From the
latter, the expansion coefficients $c_{ij}$ and $e_{ij}(m)$ are derived by
restricting $n$ to the appropriate values: $c_{ij}=J^{(1)}_{ij}$ and
$e_{ij}(m)=I^{(0)}_{ij}(m).$ Explicitly, these matrices~read\begin{eqnarray*}
c\equiv(c_{ij})&=&{\rm i}\,\frac{1024}{9!!\,\pi}\left(\begin{array}{ccc}
\sqrt{3}&\displaystyle\frac{7}{11}&\cdots\\[3ex]
\displaystyle\frac{\sqrt{15}}{11}&
\displaystyle\frac{113\,\sqrt{5}}{143}&\cdots\\[3ex]
\vdots&\vdots&\ddots\end{array}\right),\\[1ex]
e(m)\equiv(e_{ij}(m))&=&\frac{256}{7!!\,\pi\,m}\left(\begin{array}{ccc}
1&-\displaystyle\frac{1}{3\,\sqrt{3}}&\cdots\\[3ex]
-\displaystyle\frac{1}{3\,\sqrt{3}}&\displaystyle\frac{89}{99}&\cdots\\[3ex]
\vdots&\vdots&\ddots\end{array}\right),\end{eqnarray*}where$$(2\,n+1)!!\equiv
1\times 3\times\cdots\times(2\,n-1)\times(2\,n+1)\ ,\quad n=0,1,2,\dots\ .$$
The analytic results for the expansion coefficients $d_{ij}(m)$ have already
been derived~in Ref.~\cite{Lucha00:IBSEm0}; there is no need to duplicate
this formula or the matrix $d(m)\equiv(d_{ij}(m))$~here.

As has been done in Ref.~\cite{Lucha00:IBSEm0}, we shall adopt, as the
simplest model for a confining interaction between the bound-state
constituents, a linear potential: $V(r)=\lambda\,r,$ $\lambda>0.$ For this
interaction, the general expression for $V^{(\ell)}_{ij}(m)$ given in
Refs.~\cite{Lucha97,Lucha98O,Lucha98D} simplifies to (cf.\ Sec.~2.2 of
Ref.~\cite{Lucha00:IBSEm0})\begin{eqnarray*}V^{(\ell)}_{ij}(m,\lambda)&=&
\sqrt{\frac{i!\,j!}{\Gamma(2\,\ell+i+3)\,\Gamma(2\,\ell+j+3)}}\,
\frac{\lambda}{2\,m}\,\sum_{r=0}^i\,\sum_{s=0}^j\,\frac{(-1)^{r+s}}{r!\,s!}
\\[1ex]&\times&\left(\begin{array}{c}i+2\,\ell+2\\i-r\end{array}\right)
\left(\begin{array}{c}j+2\,\ell+2\\j-s\end{array}\right)\Gamma(2\,\ell+r+s+4)\
;\end{eqnarray*}the explicit potential matrices
$V^{(\ell)}(m,\lambda)\equiv\left(V^{(\ell)}_{ij}(m,\lambda)\right),$
$\ell=0,1,$ are given in Ref.~\cite{Lucha00:IBSEm0}.

The dependence of all these quantities on the dimensional parameters in the
theory, viz., the mass $m$ of the bound-state constituents and the slope
$\lambda$ of the linear potential, may be inferred already on dimensional
grounds:\begin{eqnarray*}K_{ij}(m)&=&m^2\,K_{ij}(1)\ ,\\[1ex]
b_{ij}(m)&=&m\,b_{ij}(1)\ ,\\[1ex]d_{ij}(m)&=&m\,d_{ij}(1)\ ,\\[1ex]
e_{ij}(m)&=&\frac{1}{m}\,e_{ij}(1)\
,\\[1ex]V^{(\ell)}_{ij}(m,\lambda)&=&\frac{\lambda}{m}\,V^{(\ell)}_{ij}(1,1)\
,\quad\ell=0,1,\dots\ ;\end{eqnarray*}the expansion coefficients $c_{ij}$ are
independent of the bound-state constituents' mass~$m.$ Factorizing off the
dependence on $m$ and $\lambda$ in the matrix ${\cal M}_{ij},$
Eq.~(\ref{Eq:M2-nzm}), we end up~with\begin{eqnarray*}{\cal M}_{ij}&=&
4\,m^2\,K_{ij}(1)+2\,\lambda\,\sum_{r=0}^N\left[b_{ri}(1)
+e_{ri}(1)\right]V^{(0)}_{rj}(1,1)\\[1ex]
&+&2\,\lambda\,\sum_{r=0}^N\,\sum_{s=0}^N\,c^\ast_{ri}\,V^{(1)}_{rs}(1,1)\,
d_{sj}(1)\\[1ex]
&+&\frac{\lambda^2}{m^2}\,\sum_{r=0}^N\,\sum_{s=0}^N\,\sum_{t=0}^N\,
e_{ri}(1)\,V^{(0)}_{rs}(1,1)\,e_{st}(1)\,V^{(0)}_{tj}(1,1)\\[1ex]
&+&\frac{\lambda^2}{m^2}\,\sum_{r=0}^N\,\sum_{s=0}^N\,\sum_{t=0}^N\,
c^\ast_{ri}\,V^{(1)}_{sr}(1,1)\,c_{st}\,V^{(0)}_{tj}(1,1)\ .\end{eqnarray*}
Approximate solutions of the Bethe--Salpeter equation are found by
diagonalizing~${\cal M}_{ij}.$

\section{A Few Illustrative Results}Mimicking the analysis of
Ref.~\cite{Lucha00:IBSEm0}, let us investigate first the case $d=1$ and
$N=0.$~For $i=j=0,$ we find, for the first elements of the matrices $K(1),$
$b(1),$ $c,$ $d(1),$ and $e(1),$$$K_{00}(1)=2\ ,\quad
b_{00}(1)=\frac{64}{15\,\pi}\ ,\quad c_{00}={\rm
i}\,\frac{1024}{315\,\sqrt{3}\,\pi}\ ,\quad d_{00}(1)={\rm
i}\,\frac{\sqrt{3}}{2}\ ,\quad e_{00}(1)=\frac{256}{105\,\pi}\ ,$$and, for
the expectation values $V^{(\ell)}(1,1),$ $\ell=0,1,$ of the linear potential
(cf.\ Sec.~3~of Ref.~\cite{Lucha00:IBSEm0}),$$\quad
V^{(0)}_{00}(1,1)=\frac{3}{2}\ ,\quad V^{(1)}_{00}(1,1)=\frac{5}{2}\ .$$These
matrix elements yield, for the bound-state mass~$M$ squared, the analytic
result$$M^2=8\,m^2+\frac{8896}{315\,\pi}\,\lambda
+\frac{23}{7}\left(\frac{128\,\lambda}{45\,\pi\,m}\right)^2\qquad(m\ne0)\ .$$
From this formula, we obtain, for a bound-state constituents' mass
$m=0.9\;\mbox{GeV}$~and~a typical value $\lambda=0.2\;\mbox{GeV}^2$ of the
slope of the linear potential \cite{Lucha91}, for the bound-state mass
$M=2.900\;\mbox{GeV}.$ This is only $10\%$ away from the ``exact'' result
$M=2.637\;\mbox{GeV}$ for the ground-state mass, computed for a matrix size
$d=15$ (that is, a $15\times 15$~matrix) and $N=49$ (that is, taking into
account the first 50 basis functions) in the expansions performed at
intermediate steps.

In general, that is, for matrix sizes $d>4,$ the diagonalization of our
matrix ${\cal M}_{ij}$~can be done only numerically. Table~\ref{Tab:BS-ddep}
illustrates, for the lowest-lying radial excitations,~the rather rapid
convergence of the bound-state masses $M,$ obtained as square roots of~the
eigenvalues of ${\cal M}_{ij},$ with increasing matrix size $d$ to the
numerically computed~``exact'' results.

\begin{table}[ht]\caption{Differences $M-2\,m$ of the eigenvalues $M$ of the
instantaneous Bethe--Salpeter equation and the sum of the masses $m$ of the
bound-state constituents, in units of GeV, for two spin-$\frac{1}{2}$
fermions of mass $m=0.1\;\mbox{GeV},$ experiencing an interaction
described~by a linear potential with slope $\lambda=0.2\;\mbox{GeV}^2$ and
forming bound states of radial quantum number $n_{\rm r}=0,1,2$ and
spin-parity-charge conjugation assignment $J^{PC}=0^{-+}$ (called $1^1{\rm
S}_0,$ $2^1{\rm S}_0,$ and $3^1{\rm S}_0$ in the usual spectroscopic
notation) as functions of the matrix~size $d,$ for $N=49$ (i.e., considering
50 basis vectors) in the intermediate series expansions. The last row
compares these differences with the outcome of a numerical
computation.}\label{Tab:BS-ddep}
\begin{center}\begin{tabular}{clll}\hline\hline&&\\[-1.5ex]
\multicolumn{1}{c}{$d$}&\multicolumn{1}{c}{$1^1{\rm S}_0$}&
\multicolumn{1}{c}{$2^1{\rm S}_0$}&\multicolumn{1}{c}{$3^1{\rm S}_0$}\\[1ex]
\hline\\[-1.5ex]
15&1.477&2.147&2.918\\25&1.461&2.095&2.698\\50&1.461&2.074&2.560\\[1ex]
\hline\\[-1.5ex]purely numerical&1.4613&2.0740&2.5580\\[1ex]
\hline\hline\end{tabular}\end{center}\end{table}

\begin{figure}[ht]\begin{center}\psfig{figure=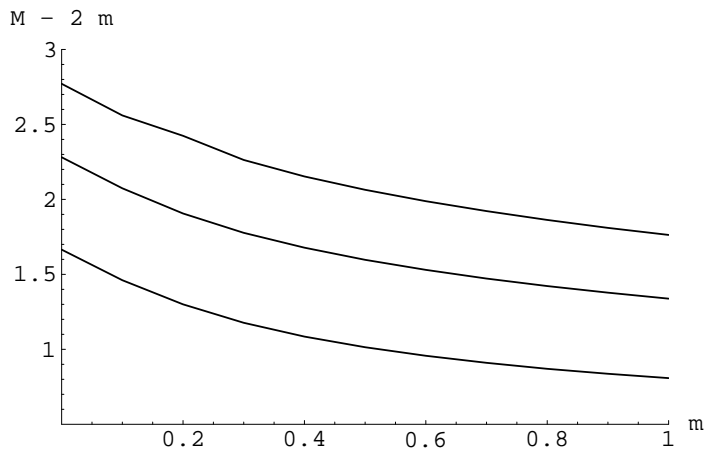,scale=0.92}
\caption{\small Dependence of the differences $M-2\,m$ of the eigenvalues $M$
of the instantaneous Bethe--Salpeter equation and the sum of the masses $m$
of the bound-state constituents for~the three lowest-lying $J^{PC}=0^{-+}$
bound states, obtained from a time-component Lorentz-vector confining
interaction kernel involving a linear potential $V(r)=\lambda\,r$ with slope
$\lambda=0.2\;\mbox{GeV}^2$, on the bound-state constituents' mass $m$ (all
masses in units of GeV). The chosen truncation parameters are $N=49$ in the
intermediate series expansions as well as $d=15$ for the matrix size (except
for $m=0.1\;\mbox{GeV},$ where, in order to increase the accuracy, $d=50$ has
been~used).}\label{Fig:differences}\end{center}\end{figure}

Figure~\ref{Fig:differences} shows the dependence of the binding energies
(i.e., the differences $M-2\,m$ of the eigenvalues $M$ of the instantaneous
Bethe--Salpeter equation and the sum of~the masses $m$ of the two bound-state
constituents) of the lowest-lying bound states on~$m.$ We observe perfect
agreement of our findings with results presented in Fig.~1 of
Ref.~\cite{Olsson96}.

\section{Summary, Conclusions, and Outlook}In the present investigation we
developed a technique for the approximative solution~of the instantaneous
Bethe--Salpeter equation with massive bound-state constituents, by a
reformulation of this equation of motion as an equivalent matrix eigenvalue
problem. Combining these findings with the analoguous result obtained within
a similar analysis for the slightly simpler case of massless bound-state
constituents presented in Ref.~\cite{Lucha00:IBSEm0}, we arrive at the
conclusion that, for a suitable choice of basis states in the Hilbert~space
of solutions, it is for a large class of interactions possible to convert the
Bethe--Salpeter equation in the instantaneous approximation for the involved
interaction kernel into~an eigenvalue problem for an explicitly known matrix,
with matrix elements given in form of analytic expressions. The main
advantage of our formalism is that, due to the scaling behaviour of the
involved quantities, in actual applications (like fitting procedures)~the
required matrices must be calculated only once (for, e.g., unit values of all
the physical parameters). As consequence of the explicit knowledge of the
matrix~representation~of the instantaneous Bethe--Salpeter equation, the
eventual diagonalization of this matrix represents the only numerical
operation required by this method of solution.

The next step must be to apply this formalism to realistic (i.e.,
phenomenologically acceptable) models of the interquark forces, capable of
reproducing the experimentally observed hadron spectra and features by
describing hadrons as bound states of quarks.

\section*{Acknowledgement}One of us (K.~M.~M.) would like to thank the Erwin
Schr\"odinger International Institute for Mathematical Physics, where part of
this work was done, for hospitality and would like to acknowledge also the
support by the NSF under grant no.~HRD-9633750.

\appendix\section{The ``Generalized Laguerre'' Basis}\label{App:Laguerre-nzm}
Our choice of basis states for $L_2(R^+)$ is fixed by the configuration-space
representation
\begin{equation}\phi_i^{(\ell)}(r)=\sqrt{\frac{(2\,m)^{2\,\ell+3}\,i!}
{\Gamma(2\,\ell+i+3)}}\,r^\ell\exp(-m\,r)\,L_i^{(2\,\ell+2)}(2\,m\,r)\ ,\quad
i=0,1,2,\dots\ ,\label{Eq:IBSE-basis-config-nzm}\end{equation}which involves
the generalized Laguerre polynomials $L_i^{(\gamma)}(x)$ (for the parameter
$\gamma$):~the latter quantities are orthogonal polynomials which are defined
by the power series \cite{Abramowitz}$$L_i^{(\gamma)}(x)=
\sum_{t=0}^i\,(-1)^t\left(\begin{array}{c}i+\gamma\\i-t\end{array}\right)
\frac{x^t}{t!}\ ,\quad i=0,1,2,\dots\ ,$$and which are orthonormalized, with
the weight function $x^\gamma\exp(-x)$, according to~\cite{Abramowitz}
$$\int\limits_0^\infty{\rm d}x\,x^\gamma\exp(-x)\,L_i^{(\gamma)}(x)\,
L_j^{(\gamma)}(x)=\frac{\Gamma(\gamma+i+1)}{i!}\,\delta_{ij}\ ,\quad
i,j=0,1,2,\dots\ .$$The necessary normalizability of the Hilbert-space basis
functions $\phi_i^{(\ell)}(r)$ is guaranteed by the positive numerical value
of the mass $m$ of the bound-state constituents: $m>0.$ The basis functions
$\phi_i^{(\ell)}(r)$ defined by Eq.~(\ref{Eq:IBSE-basis-config-nzm}) satisfy
the orthonormalization condition$$\int\limits_0^\infty{\rm d}r\,r^2\,
\phi_i^{(\ell)}(r)\,\phi_j^{(\ell)}(r)=\delta_{ij}\ ,\quad i,j=0,1,2,\dots\
.$$Note that the configuration-space representation of our basis states is
chosen to be~real. The corresponding momentum-space representation
$\phi_i^{(\ell)}(p)$ of the $L_2(R^+)$ basis states under consideration is
obtained from Eq.~(\ref{Eq:IBSE-basis-config-nzm}) by a Fourier--Bessel
transformation (recall that one is dealing here exclusively with the {\em
radial\/} parts of the $L_2(R^3)$ basis~functions):\begin{eqnarray}
\phi_i^{(\ell)}(r)&=&{\rm
i}^\ell\,\sqrt{\frac{2}{\pi}}\int\limits_0^\infty{\rm
d}p\,p^2\,j_\ell(p\,r)\,\phi_i^{(\ell)}(p)\ ,\quad i=0,1,2,\dots\
,\quad\ell=0,1,2,\dots\ ,\label{Eq:FB}\\[1ex]\phi_i^{(\ell)}(p)&=&(-{\rm
i})^\ell\,\sqrt{\frac{2}{\pi}}\int\limits_0^\infty{\rm
d}r\,r^2\,j_\ell(p\,r)\,\phi_i^{(\ell)}(r)\ ,\quad i=0,1,2,\dots\
,\quad\ell=0,1,2,\dots\ .\nonumber\end{eqnarray}Explicitly, it
reads\begin{eqnarray}\phi_i^{(\ell)}(p)
&=&\sqrt{\frac{(2\,m)^{2\,\ell+3}\,i!}{\Gamma(2\,\ell+i+3)}}\,\frac{(-{\rm
i})^\ell\,p^\ell}{2^{\ell+1/2}\,\Gamma\left(\ell+\frac{3}{2}\right)}\,
\nonumber\\[1ex]
&\times&\sum_{t=0}^i\,\frac{(-1)^t}{t!}\left(\begin{array}{c}i+2\,\ell+2\\
i-t\end{array}\right)\frac{\Gamma(2\,\ell+t+3)\,(2\,m)^t}
{(p^2+m^2)^{(2\,\ell+t+3)/2}}\nonumber\\[1ex]
&\times&F\left(\frac{2\,\ell+t+3}{2},-\frac{1+t}{2};\ell+\frac{3}{2};
\frac{p^2}{p^2+m^2}\right),\quad i=0,1,2,\dots\ ,\label{Eq:IBSE-basis-mom-nzm}
\end{eqnarray}with the hypergeometric series $F$, defined, in terms of the
gamma function $\Gamma$, by \cite{Abramowitz}
$$F(u,v;w;z)=\frac{\Gamma(w)}{\Gamma(u)\,\Gamma(v)}\,\sum_{n=0}^\infty\,
\frac{\Gamma(u+n)\,\Gamma(v+n)}{\Gamma(w+n)}\,\frac{z^n}{n!}\ .$$The
momentum-space basis functions $\phi_i^{(\ell)}(p)$ satisfy the
orthonormalization condition$$\int\limits_0^\infty{\rm
d}p\,p^2\,\phi_i^{\ast(\ell)}(p)\,\phi_j^{(\ell)}(p)=\delta_{ij}\ ,\quad
i,j=0,1,2,\dots\ .$$The availability of the Fourier transform of our basis
functions $\phi_i^{(\ell)}(r)$ in analytic form represents the main advantage
of our choice (\ref{Eq:IBSE-basis-config-nzm}). Note that the momentum-space
basis functions are real for $\ell=0,$ as well as for all even values of
$\ell$:$$\phi_i^{\ast(\ell)}(p)=\phi_i^{(\ell)}(p)\quad\mbox{for\
}\ell=0,2,4,\dots\ ,\quad\forall\ i=0,1,2,\dots\ .$$In order to get rid of
that rather difficult-to-handle hypergeometric series $F$ in
Eq.~(\ref{Eq:IBSE-basis-mom-nzm}), we occasionally take advantage of a
somewhat simplified form of the momentum-space basis functions
$\phi_i^{(\ell)}(p),$ namely, for $\ell=0$ \cite{Lucha97},\begin{eqnarray*}
\phi_i^{(0)}(p)&=&\sqrt{\frac{i!}{m\,\pi\,\Gamma(i+3)}}\,\frac{4}{p}\,
\sum_{t=0}^i\,(-2)^t\,(t+1)\left(\begin{array}{c}i+2\\i-t\end{array}\right)
\\[1ex]&\times&\left(1+\frac{p^2}{m^2}\right)^{-(t+2)/2}
\sin\left((t+2)\arctan\frac{p}{m}\right),\end{eqnarray*}and, for $\ell=1,$
\begin{eqnarray*}\phi_i^{(1)}(p)&=&-{\rm
i}\,\sqrt{\frac{m^5}{\pi\,(i+1)\,(i+2)\,(i+3)\,(i+4)}}
\,\frac{8}{p^2}\\[1ex]&\times&\sum_{t=0}^i\,\frac{(-2)^t}{t!}
\left(\begin{array}{c}i+4\\i-t\end{array}\right)
\frac{(t+3)!\,m^t}{(p^2+m^2)^{(t+3)/2}}\\[1ex]&\times&
\left[\frac{\sqrt{p^2+m^2}}{t+2}\sin\left((t+2)\arctan\frac{p}{m}\right)-
\frac{m}{t+3}\sin\left((t+3)\arctan\frac{p}{m}\right)\right].
\end{eqnarray*}The latter form of the basis functions is obtained with the
help of a suitable recursion formula~\cite{Abramowitz}.

\section{Some Useful Integrals}\label{App:integrals}With the help of the
simplified forms of the basis functions $\phi_i^{(0)}(p)$ and
$\phi_i^{(1)}(p)$ recalled in Appendix~\ref{App:Laguerre-nzm}, explicit
expressions for the integrals defined in Eq.~(\ref{Eq:integrals}) may be
found:\begin{eqnarray*}&&I^{(n)}_{ij}(m)\\[1ex]
&&=\frac{4\,m^{n-1}}{\pi\,\sqrt{(i+1)\,(i+2)\,(j+1)\,(j+2)}}\\[1ex]
&&\times\sum_{r=0}^i\,\sum_{s=0}^j\,(-2)^{r+s}
\left(\begin{array}{c}i+2\\i-r\end{array}\right)
\left(\begin{array}{c}j+2\\j-s\end{array}\right)(r+1)\,(s+1)\\[1ex]&&\times
\left[\sum_{k=0}^{|r-s|}\left(\begin{array}{c}|r-s|\\k\end{array}\right)
\frac{\Gamma(\frac{1}{2}\,(k+n+1))\,\Gamma(\frac{1}{2}\,(4+r+s+|r-s|-n-k))}
{\Gamma(\frac{1}{2}\,(5+r+s+|r-s|))}
\cos\left(\frac{k\,\pi}{2}\right)\right.\\[1ex]
&&\left.-\sum_{k=0}^{r+s+4}\left(\begin{array}{c}r+s+4\\k\end{array}\right)
\frac{\Gamma(\frac{1}{2}\,(k+n+1))\,\Gamma(\frac{1}{2}\,(8+2\,r+2\,s-n-k))}
{\Gamma(\frac{1}{2}\,(9+2\,r+2\,s))}
\cos\left(\frac{k\,\pi}{2}\right)\right]\end{eqnarray*}
and\begin{eqnarray*}&&J^{(n)}_{ij}(m)\\[1ex]&&={\rm i}\,\frac{8\,m^{n-1}}
{\pi\,\sqrt{(i+1)\,(i+2)\,(i+3)\,(i+4)\,(j+1)\,(j+2)}}\\[1ex]
&&\times\sum_{r=0}^i\,\sum_{s=0}^j\,(-2)^{r+s}\,(r+1)\,(r+2)\,(r+3)\,(s+1)
\left(\begin{array}{c}i+4\\i-r\end{array}\right)
\left(\begin{array}{c}j+2\\j-s\end{array}\right)\\[1ex]
&&\times\left\{\frac{1}{r+2}\left[\sum_{k=0}^{|r-s|}
\left(\begin{array}{c}|r-s|\\k\end{array}\right)
\frac{\Gamma(\frac{1}{2}\,(n+k))\,\Gamma(\frac{1}{2}\,(5+r+s+|r-s|-n-k))}
{\Gamma(\frac{1}{2}\,(5+r+s+|r-s|))}\right.\right.\\[1ex]
&&\times\cos\left(\frac{k\,\pi}{2}\right)\\[1ex]
&&-\left.\sum_{k=0}^{4+r+s}\,\left(\begin{array}{c}4+r+s\\k\end{array}\right)
\frac{\Gamma(\frac{1}{2}\,(n+k))\,\Gamma(\frac{1}{2}\,(9+2\,r+2\,s-n-k))}
{\Gamma(\frac{1}{2}\,(9+2\,r+2\,s))}\cos\left(\frac{k\,\pi}{2}\right)\right]
\\[1ex]&&-\frac{1}{r+3}\left[\sum_{k=0}^{|1+r-s|}\left(\begin{array}{c}
|1+r-s|\\k\end{array}\right)\right.\\[1ex]&&\times
\frac{\Gamma(\frac{1}{2}\,(n+k))\,\Gamma(\frac{1}{2}\,(6+r+s+|1+r-s|-n-k))}
{\Gamma(\frac{1}{2}\,(6+r+s+|1+r-s|))}\cos\left(\frac{k\,\pi}{2}\right)
\\[1ex]&&\left.\left.-\sum_{k=0}^{5+r+s}
\left(\begin{array}{c}5+r+s\\k\end{array}\right)
\frac{\Gamma(\frac{1}{2}\,(n+k))\,\Gamma(\frac{1}{2}\,(11+2\,r+2\,s-n-k))}
{\Gamma(\frac{1}{2}\,(11+2\,r+2\,s))}\cos\left(\frac{k\,\pi}{2}\right)\right]
\right\}.\end{eqnarray*}

\end{document}